\pdfoutput=1
\documentclass[a4paper,11pt]{article}
\usepackage{tikz-cd}
\usepackage[dvipsnames]{xcolor}
\usepackage{amsmath,amssymb,amsthm,amsfonts,mathrsfs}
\usepackage{bm, cancel, ulem}
\usepackage{enumitem}
\usepackage{booktabs, comment, tikz}
\usepackage{tabularx}
\usepackage{jheppub}
\usepackage[T1]{fontenc}
\usepackage[utf8]{inputenc}
\usepackage{hyperref}
\usetikzlibrary{calc}
\hypersetup{
	linktoc=all,
	colorlinks=true,
	linkcolor=blue,
	citecolor=magenta,
	urlcolor=red
}

\newcommand{\cU}{\mathcal U}
\newcommand{\cV}{\mathcal V}
\newcommand{\cW}{\mathcal W}

\newcommand{\cL}{\mathcal{L}}

\usepackage[most]{tcolorbox}

\tcbset{highlight math style={left=02mm,right=02mm,top=02mm,bottom=02mm}} 
\usepackage{empheq}

\makeatletter \@addtoreset{equation}{section}

\newcommand\bc{\begin{center}}
	\newcommand\ec{\end{center}}

\begin{document}
	
	\vspace*{-5cm}
	\title{\bc
		\huge{Carrollian Worldsheets on Carrollian Targets}\\
		\centerline{\LARGE{When does ILST describe a null string theory?}}
        \ec}
	\author[a]{M.M. Sheikh-Jabbari }
	\author[b]{, H. Yavartanoo}
	
	\affiliation[a]{School of Physics, Institute for Research in Fundamental Sciences (IPM),
		P.O. Box 19395-5531, Tehran, Iran}
	\affiliation[b]{Beijing Institute of Mathematical Sciences and Applications (BIMSA),
		Huairou District, Beijing 101408, P. R. China}
	
	\emailAdd{jabbari@theory.ipm.ac.ir}
	\emailAdd{yavar@bimsa.cn}

\abstract{Null string theory (NST) is a two-dimensional ($2d$) $\sigma$-model on  a Carrollian worldsheet with $2d$ diffeomorphisms, Weyl and Carroll-Weyl (CW) scalings  as gauge symmetries whereas the Isberg–Lindstr\"om–Sundborg–Theodoridis (ILST) theory on a generic target space only realizes the first two and not the CW gauge symmetry. We show in this work that NST on the Carrollian target spaces (CTS) is well-defined if the background admits a homothetic kernel generator and that, it is described by the ILST theory on the same CTS. To this end, we formulate a minimal null string theory on CTS, starting from first principles and then discuss beyond minimal theories as deformations away from minimality. We study some classes of such deformations involving the clock one-from for ruled Carrollian spaces  and/or connection on the Carrollian space. We discuss implications of this for the recently introduced null string holography. As examples CTS, we  briefly discuss null strings probing black hole horizons and the null string holography in this context.}

	\maketitle

	\section{Introduction}
	\label{sec:intro}

In physics, as in real world, we deal with Lorentzian spacetimes, with a time direction and a metric which is not positive (semi)definite; it allows for spacelike, timelike and null, in our conventions respectively associated with positive, negative and zero, separations. Due to analyticity behavior of physical observables (e.g. in a quantum field theory),  to perform the computations and obtain physical results it is customary to Wick-rotate and work in Euclidean spaces, with positive-definite norm. $D$ dimensional Lorentzian manifolds, however, allow  allow for codimension-1 null hypersurfaces whose induced metric has rank $D-2$. These are null surfaces in these manifolds and are physically very important, because they locally behave as ``horizons''. Physics on null surfaces is relevant in several physically interesting cases, like black holes, cosmology and asymptotics of flat space. It is desirable to formulate physics on null surfaces on their own and not just as boundaries of a given Lorentzian spacetime. 

The geometry associated with null surfaces is call \textit{Carrollian geometries} \cite{Duval:2014uva, Duval:2014uoa}, which are neither Lorentzian nor Euclidean; these are a third class of manifolds, ``zero signature manifolds'' \cite{Henneaux:1979vn} and deserve to be studied on their own turn, aside from embedding in a one higher-dimension Lorentzian manifold, see \cite{Ciambelli:2025unn} for a recent review. As mentioned, a $D$ dimensional Carrollian manifold has a degenerate metric with one degenerate direction and $D-1$ dimensional positive-definite metric. So, to minimally describe the manifold we need a degenerate metric $h_{\mu\nu}$ and a kernel vector along its zero eigenvector direction $V^\mu$; hence minimally a Carrollian manifold is described by $(h, V)$ pair, such that $h_{\mu\nu} V^\nu=0$. We will give a concise review of Carrollian geometries in section \ref{sec:review}.

In this work we study a problem that involves two Carrollian geometries, a $2d$ Carrollian worldsheets and $D$ dimensional Carrollian target space (CTS). Strings with Carrollian worldsheets, null strings, has half a century literature, since \cite{Schild:1976vq}. In early 1990s the Isberg–Lindstr\"om–Sundborg–Theodoridis (ILST)  work \cite{Isberg:1992ia, Isberg:1993av} it was put in a more rigorous mathematical setting and received a renewed attention by the surge in exploring Carrollian physics in the last 5-6 year, see e.g. \cite{Bagchi:2025vri, Ruzziconi:2026bix} and references therein. In particular, recently, based on some observations \cite{Sheikh-Jabbari:2026cnj, Sheikh-Jabbari:2026tpf, Sheikh-Jabbari:2026vqh} (see however, \cite{Lindstrom:2026tua, Lindstrom:2026zno, Lindstrom:2026quz}, we gave a first principle definition of a null string theory \cite{Sheikh-Jabbari:2026kwr}. This definition is based on the fact that string theory is a $2d$ $\sigma$-model in which all geometric worldsheet symmetries are gauged, this includes the ``Carroll-Weyl (CW) scaling'' that is particular to Carrollian worldsheets, besides the standard $2d$ diffeomorphisms and Weyl scaling (of the Polyakov tensile strings \cite{Polyakov:1981rd}).\footnote{The CW gauged worldsheet theory, was considered under ``conformal null string'' theory in \cite{Gustafsson:1994kr}.} This definition was then extended to generic Lorentzian target spaces that {admit at least a conformal Killing vector required by the CW gauge symmetry}  \cite{Sheikh-Jabbari:2026rcr}.

The spacetime background that strings probe/live in, the target space, is supposed to be a dynamical object that is formed and sensed by the string probing it \cite{Green:1987sp, Polchinski:1998rq}. Nonetheless, the signature of the target space, being Lorentzian, Euclidean or Carrollian, does not seem to be a dynamical quantity and is chosen from the outset. The Carrollian target spaces, while mentioned in the literature \cite{Bagchi:2022iqb, Bagchi:2023cfp, Bagchi:2024rje, Harksen:2024bnh, Figueroa-OFarrill:2025njv, Bagchi:2026wcu, Bagchi:2026qpi}, have not received due attention. In particular, as pointed out in \cite{Bagchi:2023cfp, Bagchi:2024rje}, strings probing black hole horizons are expected to also have Carrollian worldsheets. 

In this work we focus on formulating null string theory (as defined in \cite{Sheikh-Jabbari:2026kwr}) on Carrollian target spaces (CTS) or ruled Carrollian target spaces (RCTS). We show that for these cases, the ILST theory and the null string theory are (classically) equivalent. 

The paper is organized as follows. In section~\ref{sec:review} we set the stage by reviewing basics of Carrollian geometries and the null string theory. In section \ref{sec:NSTonCTS} we discuss minimal null string theory (NST) and ILST theories on CTS and show that the two are equivalent. In section \ref{sec:NST-RCTS} we extended the discussion to the ruled Carrollian target spaces (RCTS) and show again the minimal NST and ILST theories on RCTS are equivalent. In section \ref{sec:examples} we discuss some examples with (R)CTS and discuss the equivalence established in previous sections. Section \ref{sec:discussion} contains our concluding remarks and outlook. In appendix \ref{sec:hamil}  we have gathered some of technical analyses and in appendix \ref{app:tensile-CTS} we discuss tensile strings on Carrollian target spaces.

\paragraph{Conventions.} In this work,  ``null string theory'' (NST) refers to a theory in which $2d$ diffeomorphisms, Weyl and CW scalings are gauged, and when CW scaling is not gauged, it will be called ILST theory. 

\section{Review of the background material}\label{sec:review}

This work concerns Carrollian worldsheets on Carrollian target spaces. Hence, it is appropriate to review basics of Carrollian geometry and manifolds, and also what null string theory is, setting the stage for the subsequent sections. 

\subsection{Basics of Carrollian geometry}\label{sec:Carrollian-geometry-review}

Carrollian manifolds are ``signature zero'' geometries \cite{Henneaux:1979vn} that describe geometry of null surfaces. A $D$-dimensional manifold $\mathcal C$ carries a minimal Carrollian structure when equipped with a pair $(h_{\mu\nu},V^\mu)$, where $h_{\mu\nu}$ acts as a smooth, positive-semidefinite symmetric metric of constant rank $D-1$ with signature
\begin{equation}\label{metric-signature}
(0,+,\ldots,+),
\end{equation}
and the nowhere-vanishing vector field $V^\mu$ spans its one-dimensional kernel,
\begin{equation}
h_{\mu\nu}V^\nu=0.
\label{eq:minimalCarroll}
\end{equation}
Integral curves of $V^\mu$ establish the distinguished Carrollian direction. The tensor $h_{\mu\nu}$ supplies a strictly positive-definite metric for the $D-1$ transverse directions. We enforce no invariance condition on $h_{\mu\nu}$ along $V^\mu$. The local expansion and shear operators track the exact metric deformation along the kernel flow.

\paragraph{Ruled Carrollian geometries.} These are manifolds equipped with an additional one-form $U_\mu$. The required normalization is
\begin{equation}
U_\mu V^\mu=1.
\label{eq:UVnormalization}
\end{equation}
The full geometric data set becomes the triplet
\begin{equation}
(h_{\mu\nu},V^\mu,U_\mu).
\end{equation}
If the kernel foliation forms a fibration, $\ker U$ identifies a horizontal distribution. The field $U$ then acts as an Ehresmann connection. The metric-kernel pair $(h,V)$ leaves this covector undetermined. The one-form $U_\mu$ represents strictly independent boundary data. The inclusion of $U_\mu$ uniquely determines the symmetric transverse inverse $h^{\mu\nu}$ through the algebraic relations
\begin{equation}
h^{\mu\nu}U_\nu=0,
\qquad
h^{\mu\alpha}h_{\alpha\nu}
+V^\mu U_\nu =
\delta^\mu{}_\nu.
\label{eq:CarrollCompleteness}
\end{equation}
The tensor $h^{\mu\nu}$ is not a true inverse on the full tangent space. It inverts $h_{\mu\nu}$ solely on the transverse subbundle annihilated by $U_\mu$.

Embedded null hypersurfaces in Lorentzian manifolds naturally realize Carrollian geometries. The induced metric is degenerate. The null normal generates its kernel. Selecting an ambient transverse rigging vector projects out a normalized clock one-form, completing a ruled Carrollian structure. The present formulation remains entirely intrinsic and relies on no ambient spacetime embedding. See e.g \cite{Ciambelli:2025unn}, section~5.1, for more details.

\paragraph{Scalar clocks.} In defining the clock one-form we did not specify $\cL_V U$ and it need not be an exact one-form. When it is exact, $\cL_V U=0$ and one can write $U=\text{d}\varphi$. For this case,  $V(\varphi)=V^\mu\partial_\mu\varphi=1$. In these cases the scalar may be used to define the clock. In the examples we discuss in section \ref{sec:NST-RCTS} examples involving such scalars, which play the role of a generic dilaton field. The exponential of these scalars, $\beta=e^{2\varphi}$ (with $V(\beta)=2\beta$) appear as couplings in the null string worldsheet theory.

\paragraph{Carrollian shear.} Metric evolution along the kernel direction is governed by the Lie derivative,
\begin{equation}
\theta_{\mu\nu} := \frac12 \mathcal L_V h_{\mu\nu},
\qquad
\theta_{\mu\nu}V^\nu = 0 .
\label{eq:expansionTensor}
\end{equation}
The deformation tensor $\theta_{\mu\nu}$ is symmetric. The transversality condition $\theta_{\mu\nu}V^\nu = 0$ follows algebraically from $h_{\mu\nu}V^\nu = 0$ combined with $\mathcal L_V V = 0$.  We split the deformation tensor into pure trace and trace-free shear components,
\begin{equation}
\theta_{\mu\nu} = \frac{\Theta}{D-1}h_{\mu\nu} + {\cal S}_{\mu\nu}, \qquad \mathcal L_Vh_{\mu\nu} = 2\frac{\Theta}{D-1}h_{\mu\nu} + 2{\cal S}_{\mu\nu},
\label{eq:deformationDecomposition}
\end{equation}
where $D$ is the dimension of the Carrollian geometry. The scalar $\Theta$ tracks isotropic variations of the transverse metric along the kernel vector. The shear tensor ${\cal S}_{\mu\nu}$ captures area-preserving deformations. A vanishing expansion $\Theta=0$ does not enforce $\mathcal L_Vh_{\mu\nu}=0$. The metric can undergo pure shear evolution.

The shear-free condition,
\begin{equation}
{\cal S}_{\mu\nu} = 0,
\end{equation}
yields the conformal Killing equation for kernel vector $V$,
\begin{equation}
\mathcal L_V h_{\mu\nu} = 2\omega h_{\mu\nu},
\qquad
\omega = \frac{\Theta}{D-1},
\label{eq:shearFreeReview}
\end{equation}
for some scalar field $\omega$. The flow acts as a local conformal transformation of the transverse space whenever $\omega \neq 0$. The flow is a transverse isometry when $\omega = 0$.

\paragraph{The two Carrollian scalings.} As discussed in \cite{Sheikh-Jabbari:2026vqh}, for Carrollian geometries one can consider two independent scalings. While we mainly discuss these two scalings for worldsheet of null strings, where they appear as gauge symmetries, similar scalings are also there for Carrollian target spaces, though not as a gauge symmetry; see next section for more analyses. 

The two scalings are associated with conformal rescaling of the metric and a rescaling of the kernel generator,
\begin{equation}\label{two-target-scalings}
\widetilde h=e^{2F}h,\qquad \widetilde V=fV,
\end{equation}
with \(f\) nowhere vanishing.  For the ruled Carrollian geometries, \eqref{two-target-scalings} should be supplemented with $U_\mu\to f^{-1}U_\mu$. Under this rescaling, 
\begin{equation}
  \widetilde{\Theta}= f \big(\Theta+ (D-1)V(F)\big),\qquad  \widetilde{\cal S}_{\mu\nu}=f e^{2F}{\cal S}_{\mu\nu}.
\end{equation}
As we see one can't set the shear to zero using the two scalings and a shear-free case is mapped to a shear-free case. Whereas, a non-zero expansion can be set to zero upon an appropriate choice of $F$. Therefore, within a given conformal class and kernel line distribution, a non-vanishing ${\cal S}_{\mu\nu}$ strictly forbids the existence of a homothetic generator. So, if $V$ is a homothetic Killing vector of $h$, $\cL_V h_{\mu\nu}=2h_{\mu\nu}$, then it can be isometry of 
$\widetilde h$,  $\mathcal L_V \widetilde h=0$, if $V^\mu\partial_\mu F=-1$. This equation always has local solutions because $V$ has no fixed points. Global solutions depend on the topology of the integral curves. The vanishing of the shear tensor is necessary and sufficient for the kernel flow to stabilize the transverse conformal class. 

\paragraph{Connection compatibility in $(h,V)$ Carrollian manifolds.}
The data \((h,V)\) or \((h,V,U)\) do not by themselves specify an affine
connection \(\nabla_\mu\). For a nondegenerate metric, metric compatibility
together with vanishing torsion uniquely determines the Levi-Civita
connection given by the Christoffel symbols $\Gamma^\rho{}_{\mu\nu}$. In the Carrollian case, compatibility with the chosen pair \((h,V)\) may be imposed through
\begin{equation}
 \nabla_\mu h_{\alpha\beta}=0,\qquad
 \nabla_\mu V^\nu=0.
 \label{eq:Carroll-connection-compatibility}
\end{equation}
Metricity alone preserves the kernel line.\footnote{We note that $\nabla_\mu h_{\alpha\beta}=0$ together with $V^\mu h_{\mu\beta}=0$ implies that $  h_{\alpha\beta}\nabla_\mu V^\alpha=0$. So, $\nabla_\mu V^\alpha=0$ is not implied by $\nabla_\mu h_{\alpha\beta}=0$ and imposes further restrictions on the connection. } The second condition preserves
the chosen kernel vector as well; it does not follow from the first.
These conditions generally leave freedom in the connection, even when
a ruling \(U_\mu\) is supplied.

Let $C^\rho_{\mu\nu}$ be the connection and \(T^\rho_{\mu\nu}:=2C^\rho_{[\mu\nu]}\) its antisymmetric part, the torsion. Then, 
\begin{equation} \label{eq:RCTS-torsion-expansion}
\begin{split}
 2\theta_{\mu\nu}=\mathcal L_Vh_{\mu\nu}
 &= V^\alpha\nabla_\alpha h_{\mu\nu}+ h_{\mu\alpha}\nabla_\nu V^\alpha + h_{\alpha\nu}\nabla_\mu V^\alpha + V^\rho T^\lambda{}_{\rho\mu}h_{\lambda\nu}
 +V^\rho T^\lambda{}_{\rho\nu}h_{\mu\lambda}\\
&=V^\alpha (\nabla_\alpha h_{\mu\nu}-\nabla_\nu h_{\mu\alpha} - \nabla_\mu h_{\alpha\nu}) + V^\rho T^\lambda{}_{\rho\mu}h_{\lambda\nu}
 +V^\rho T^\lambda{}_{\rho\nu}h_{\mu\lambda}.        
\end{split}
\end{equation}
Therefore, torsion-free + metricity $\nabla_\mu h_{\alpha\beta}=0$ imply \(\theta_{\mu\nu}=0\) or vanishing of expansion and shear.
Conversely,  \(\theta_{\mu\nu}\neq0\), including the homothetic case
\(\mathcal L_Vh=2h\), implies nonzero torsion or nonmetricity \cite{Ciambelli:2025unn}, and finally,
\begin{equation}
    \begin{split}
        \text{Homotheticity} \ \mathcal L_Vh=2h\ + \text{metricity}:& \qquad\quad 2h_{\mu\nu} =V^\rho T^\lambda{}_{\rho\mu}h_{\lambda\nu} +V^\rho T^\lambda{}_{\rho\nu}h_{\mu\lambda}\\
        \text{Homotheticity} \ \mathcal L_Vh=2h\ + \text{torsion-free}:& \quad h_{\mu\nu} =\frac12V^\alpha (\partial_\alpha h_{\mu\nu}-\partial_\nu h_{\mu\alpha} - \partial_\mu h_{\alpha\nu}):=- V^\alpha \Gamma_{\mu\nu}{}_{\ \alpha}
    \end{split}
\end{equation}
 
{Therefore, on the homothetic backgrounds considered below, any target-space affine connection must have nonzero torsion or nonmetricity (cf. discussion of the next section), and backgrounds with $\mathcal L_Vh=0$ do not have this obstruction.} 

\paragraph{Kernel flow invariant connections.} As another specific choice of connection is when connection does not change along the flow of the kernel $V^\mu$. While connection $C_{\mu\nu}^\alpha$ is not a tensor its Lie derivative is:
\begin{equation}\label{connection-V-flow}
    \cL_V C_{\mu\nu}^\alpha= \nabla_\mu (\nabla_\nu V^\alpha+ T_{\beta\nu}^\alpha V^\beta)+ R^\alpha{}_{ \nu\beta\mu} V^\beta,
\end{equation}
where $R^\alpha{}_{ \nu\beta\mu}$ is the Riemann curvature. 
Vanishing of the above is guaranteed if each of the terms in the right-hand-side vanish. 

However, we note that $\cL_V C_{\mu\nu}^\alpha=0$ does not necessarily imply this. For example, in adapted coordinates with $V=\partial_u$, this condition simply means that the connection coefficients are independent of $u$: {($\partial_u C^\rho_{\mu\nu}=0$, which does not require $V^\mu C^\rho_{\mu\nu}$ to vanish, although its $u$-derivative does vanish in these adapted coordinates.)} If we supplement \eqref{connection-V-flow} with compatibility condition $\nabla_\nu V^\alpha=0$, we learn that $(\nabla_\mu  T_{\beta\nu}^\alpha + R^\alpha{}_{ \nu\beta\mu}) V^\beta=0$. 

The minimal null-string action does not require choosing an independent affine connection. However, it is natural to leave this choice open and to specify additional
connection data when introducing connection-dependent couplings in
section \ref{sec:NST-CTS-affine}. This target affine connection is distinct
from the Ehresmann clock \(U_\mu\) and the worldsheet CW connection
\(\mathcal W_a\) introduced below. 

\subsection{Null string theory: a quick review}
\label{sec:NST-review}

In the null string theory, the worldsheet is a $2d$ Carrollian geometry spanned by coordinates $\sigma^a=(\tau,\sigma)$. The worldsheet kernel vector $\mathcal V^a$ is nowhere vanishing and normalized such that it has weight $1/2$. We take the target space to be spanned by coordinates $X^\mu$. In our analysis we discuss two class of actions on the Carrollian worldsheet, the ILST theory \cite{Isberg:1993av} and the null string theory defined in \cite{Sheikh-Jabbari:2026kwr}. 

\paragraph{ILST kinematics.} We define the ILST action for  classical tensionless of Polyakov action for tensile string on a nondegenerate Lorentzian target $g_{\mu\nu}$ as,
\begin{equation}
S_{\mathrm{ILST}}[g;\mathcal V]
=\frac{\kappa}{2}\int d^2\sigma\,
\mathcal V^a\mathcal V^b g_{\mu\nu}(X)
\partial_aX^\mu\partial_bX^\nu
=\frac{\kappa}{2}\int d^2\sigma\,g_{\mu\nu} P^\mu P^\nu,\qquad P^\mu:=\cV^a\partial_a X^\mu, 
\label{eq:NSTreview_ILST}
\end{equation}
where $\kappa$ supplies an overall normalization constant rather than a physical string tension. Note that the ILST theory \cite{Isberg:1993av} is the above action for $g_{\mu\nu}=\eta_{\mu\nu}$, the Minkowski metric. We vary $\mathcal V^a$ to extract the field equations
\begin{equation}
P_\mu\partial_aX^\mu=0,
\qquad
\gamma_{ab}\mathcal V^b=0,
\qquad
\gamma_{ab}:=g_{\mu\nu}\partial_aX^\mu\partial_bX^\nu,\qquad P_\mu:=g_{\mu\nu} P^\mu
\label{eq:NSTreview_nullness}
\end{equation}

The induced worldsheet metric $\gamma_{ab}$ is strictly degenerate on any patch with $\mathcal V^a\neq0$, in other words, on-shell, $(\gamma_{ab}, \cV^a)$ defines the Carrollian worldsheet geometry. The vector $P^\mu$ serves as the non-vanishing null tangent on an immersed branch. 

The action \eqref{eq:NSTreview_ILST} is invariant under $2d$ diffeomorphisms (generated by $\xi^a$) plus the Weyl scaling (generated by $\rho$) \cite{Sheikh-Jabbari:2026vqh}, 
\begin{equation}
    \begin{split}
        \delta_{\xi, \rho} X^\mu= \xi^a\partial_a X^\mu, \qquad \delta_{\xi,\rho} \cV^a =\xi^b\partial_b\cV^a- \cV^b\partial_b\xi^a +\frac12 (\partial_b\xi^b)\cV^a, \qquad \delta_{\xi,\rho} {\gamma_{ab}}={\cal L}_\xi\gamma_{ab}
    \end{split}
\end{equation}
(That is, $X^\mu, \cV^a, \gamma_{ab}$ are all invariant under Weyl scaling $\rho$.)

We select a local gauge (partially fixing $2d$ diffeomorphisms) $\mathcal V^a=(1,0)$, see e.g. \cite{Sheikh-Jabbari:2026cnj} for more details. The resulting embedding constraints force the worldsheet to describe a congruence of affinely parametrized null geodesics subject to the condition $g_{\mu\nu}\dot X^\mu X'^\nu=0$.

\paragraph{Carroll--Weyl (CW) completion.} The minimal null string theory is an extension of the ILST theory which also realizes the Carroll-Weyl (CW) scaling \cite{Sheikh-Jabbari:2026kwr} generated by $\chi=\chi(\tau, \sigma)$, under which,
\begin{equation}
    \delta_\chi\cV^a=-\chi \cV^a,\qquad {\delta_\chi X^\mu= \chi K^\mu(X).}
\end{equation}
To realize the above as a gauge symmetry, 
\begin{enumerate}
    \item the target spacetime must have a homothety vector $K^\mu$, $\mathcal L_Kg=2g$ \cite{Sheikh-Jabbari:2026rcr}; 
    \item the $\partial_a$ should be replaced with a CW covariant derivative \cite{Sheikh-Jabbari:2026kwr},
\begin{equation}
D_aX^\mu:=\partial_aX^\mu+\mathcal W_aK^\mu(X)\,,\qquad \delta_\chi\mathcal W_a=-\partial_a\chi.
\end{equation}
and therefore, $\delta_\chi(D_aX^\mu)=\chi\partial_\nu K^\mu D_aX^\nu$. 
\end{enumerate} 
The minimal null string theory (NST) is then defined by the action,
\footnote{We note that,
$\delta_\chi S_{\mathrm{ILST}}
=\kappa\int d^2\sigma\,
\mathcal V^a\partial_a\chi\,g_{\mu\nu}K^\mu P^\nu.
$ This integral remains nonzero off shell for an arbitrary parameter $\chi$, while it vanishes for the subset of CW scaling satisfying $V^a\partial_a\chi=0$. See \cite{Sheikh-Jabbari:2026cnj} for more discussions.}
\begin{equation}
S_{\mathrm{NST}}[g, K;\mathcal V, \cW]=\frac{\kappa}{2}\int d^2\sigma\,
\mathcal V^a\mathcal V^b g_{\mu\nu}D_aX^\mu D_bX^\nu.
\label{eq:NSTreview_CWaction}
\end{equation}
We note that the CW connection (gauge field) $\cW_a$ is not the worldsheet Ehresmann covector and the Carrollian worldsheet is not necessarily ruled.

\paragraph{Hamiltonian constraints.} The three gauge transformations can be fixed and dealt with, see e.g. \cite{Sheikh-Jabbari:2026tpf}, provided one imposes the constraints
\begin{equation}\label{eq:NSTreview_constraints}
    {\cal C}_a=\Pi_\mu\partial_a X^\mu\approx0,\qquad
\mathcal C_3=K^\mu\Pi_\mu\approx0.
\end{equation}
where
\begin{equation}
    \Pi_\mu=\frac{\partial\mathscr L}{\partial P^\mu}=\kappa g_{\mu\nu}\bigl(P^\nu+wK^\nu\bigr),\qquad w:=\mathcal V^a\mathcal W_a
\end{equation}
is the canonical momentum for the CW-gauged action. The homothety supports closure of this canonical algebra.
The specified homothety condition guarantees these constraints form a closed first-class algebra. We set $g_{\mu\nu}=\eta_{\mu\nu}$ and $K^\mu=X^\mu$ to recover the standard flat-space constraint $X\cdot P\approx0$
; this specific geometric choice pins down an origin for the homothety \cite{Sheikh-Jabbari:2026cnj, Sheikh-Jabbari:2026tpf}.

\section{NST on shear-free CTS, the minimal setup}\label{sec:NSTonCTS}

In the previous section we reviewed the minimal NST and ILST theory on non-degenerate target spaces. One can readily observe that for shear-free Carrollian target spaces (CTS), some of the steps in the previous section should be revisited. In this section we study and construct both of the theories on CTS. In this section we establish that the minimal NST with CTS is described by the ILST action on the same CTS. 

\subsection{ILST theory on an unruled CTS}\label{sec:CTS-ILST}

Consider a $D$-dimensional ($D\geq2$) CTS $\mathcal C$ $(h, V)$. 
A direct (and minimal) generalization of discussion in section \ref{sec:NST-review}, eq.\eqref{eq:NSTreview_ILST}, to the Carrollian target case is,
\begin{equation}
S_{\rm ILST}[(h,V);\mathcal V]
=\frac{\kappa}{2}\int d^2\sigma\,
h_{\mu\nu}(X)P^\mu P^\nu\,,
\qquad P^\mu:=\mathcal V^a\partial_aX^\mu.
\label{eq:S0}
\end{equation}
This notation explicitly tracks the target structure $(h,V)$. The vector $V$ enters the action implicitly through the kernel condition $h_{\mu\nu} V^\nu=0$. The theory requires no inverse of the degenerate target metric and hence no need for the kernel covector on the CTS. Moreover, the theory does not require $V$ to be shear-free. 

Since $\mathcal V^a$ has weight 1/2,  the integrand into a scalar density of weight one and the action \eqref{eq:S0} preserves worldsheet diffeomorphism invariance. Worldsheet Weyl scaling is a gauge symmetry of \eqref{eq:S0} without the need for any extra gauge field/connection, because the Weyl scaling  acts trivially on both $X^\mu$ and $\mathcal V^a$ as established in section \ref{sec:NST-review}. Independent CW scaling must be verified off shell and $h_{\mu\nu} V^\nu=0$ is not sufficient to guarantee CW gauge invariance.

\subsection{Minimal NST and  decoupling of the CW connection}\label{sec:NSTonCTS-def}

The minimal NST action of section \ref{sec:NST-review} generalizes to the CTS, as it does not require the inverse of target metric. Let $K^\mu(X)$ be a homothety vector field of the CTS $(h, V)$, $\mathcal L_Kh=2h$, and consider the CW scaling supplemented with a target space diffeomorphism, 
\begin{equation}
\delta_\chi X^\mu=\chi K^\mu(X),
\qquad
\delta_\chi\mathcal V^a=-\chi\mathcal V^a,
\label{eq:CWtrans}
\end{equation}
where $\chi(\sigma^a)$ is an arbitrary parameter. The homothety  vector $K$ is  distinct from the kernel vector $V$ and $V$ is not necessarily shear-free, yet. We introduce a worldsheet one-form $\mathcal W_a$ and construct the covariant derivative
\begin{equation}
D_aX^\mu=\partial_aX^\mu+\mathcal W_aK^\mu(X),
\qquad
\delta_\chi\mathcal W_a=-\partial_a\chi.
\label{eq:covariant}
\end{equation}
These rules combine with \eqref{eq:CWtrans} to give
\begin{equation}
\delta_\chi(D_aX^\mu)
=\chi\partial_\nu K^\mu D_aX^\nu.
\label{eq:Dhomogeneous}
\end{equation}
The gauged functional
\begin{equation}
S_{\rm NST}[(h,V),K;\mathcal V,\mathcal W]
=\frac{\kappa}{2}\int d^2\sigma\,
\mathcal V^a\mathcal V^b h_{\mu\nu}(X)
D_aX^\mu D_bX^\nu
\label{eq:SCW}
\end{equation}
is exactly invariant under local CW transformations, since a straightforward calculation yields
\begin{equation}
\delta_\chi S_{\rm NST}
=\frac{\kappa}{2}\int d^2\sigma\,
\chi\mathcal V^a\mathcal V^b
(\mathcal L_Kh-2h)_{\mu\nu}D_aX^\mu D_bX^\nu=0,
\label{eq:CTS-NSTvariation}
\end{equation}
which is guaranteed if $\mathcal L_Kh=2h$.

\subsection{Equivalence of NST and ILST on shear-free homothetic CTS}\label{sec:cwvar}

The definition of NST and \eqref{eq:CTS-NSTvariation} does not require the homothety to be along the kernel. Alignment with the kernel is a separate physical condition. If we assume the alignment, namely $h_{\mu\nu} K^\nu=0$, 
\begin{equation}
\begin{split}
h_{\mu\nu}D_aX^\mu D_bX^\nu
&={}h_{\mu\nu}\partial_aX^\mu\partial_bX^\nu
+\mathcal W_a h_{\mu\nu}K^\nu\partial_bX^\mu+\mathcal W_b h_{\mu\nu}K^\nu \partial_aX^\mu 
+\mathcal W_a\mathcal W_b h_{\mu\nu}K^\mu K^\nu
\\
&={}h_{\mu\nu}\partial_aX^\mu\partial_bX^\nu.
\end{split}
\label{eq:decoupling}
\end{equation}
Therefore, 
\begin{equation}
S_{\rm NST}[(h,V),K;\mathcal V,\mathcal W]
=S_{\rm ILST}[(h,V);\mathcal V],
\label{eq:SCWeq}
\end{equation}
and assuming that the left and right hand sides are individually well-defined, implies that
\begin{equation}
h_{\mu\nu}K^\nu=0,
\qquad \mathcal L_Kh_{\mu\nu}=2h_{\mu\nu}.
\label{eq:CWconditions}
\end{equation}
This is an off-shell identity and relies neither on embedding equations nor on gauge-fixing $\mathcal W_a=0$. Demanding equality of these functionals across arbitrary fields and arbitrary $\mathcal W_a$ reverses the logic; Equality \eqref{eq:SCWeq} requires kernel alignment; homothety (which is needed for the LHS to be well-defined)  ensures CW invariance.

Some comments are in order:
\begin{enumerate}[leftmargin=12pt, itemsep=-2pt]
\vspace*{-4mm}
    \item Since the target has a one-dimensional kernel $V$, any nonvanishing CW generator satisfying \eqref{eq:CWconditions} should be of the form $K=fV$ on the local patch. 
    \item We note that ${\cal L}_V h_{\mu\nu}\neq 0$ and ${\cal L}_V h_{\mu\nu}=\frac{2}{f} h_{\mu\nu}$. Therefore, $(h, V)$  should be shear-free. 
    \item The information about $K$ (or equivalently $f$) is already contained in $(h, V)$ pair; $f$ is not an independent function.
    \item Eq.~\eqref{eq:SCWeq} also implies that the ILST on a CTS describes a NST as defined in previous subsection  only when the Lie derivative of the Carrollian metric satisfies ${\cal L}_V h_{\mu\nu}=\frac{2}{f} h_{\mu\nu}$, for a {smooth, nowhere-vanishing and finite}  arbitrary function $f$. 
\end{enumerate}

The connection equation on CTS $(h, V)$, degenerates to the trivial identity for 
\begin{equation}
\frac{\delta S_{\rm NST}}{\delta\mathcal W_a}\equiv0,
\label{eq:noAeq}
\end{equation}
allowing arbitrary additive shifts in the decoupled action:
\begin{equation}
\delta_\alpha\mathcal W_a=\alpha_a(\tau,\sigma).
\label{eq:Ashift}
\end{equation}
We quotient out this redundant sector locally. Assuming no independent boundary couplings for the connection, this reduction leaves exactly the ILST theory and its intrinsic CW symmetry. The connection generates no secondary Gauss-law equation.

To complete the discussion, let us ask the reverse question: when is an ILST theory \eqref{eq:S0} describing a NST? i.e. when is the ILST action  invariant under \textit{off-shell} transformations \eqref{eq:CWtrans}? We recall that the $X^\mu$ transformation in \eqref{eq:CWtrans} is a diffeomorphism on the target space and therefore,  

\begin{equation}
\begin{split}
\delta_\chi S_{\rm ILST}
=&\frac{\kappa}{2}\int d^2\sigma\,\left[
\chi(\mathcal L_Kh-2h)_{\mu\nu}P^\mu P^\nu +2
(\mathcal V^a\partial_a\chi)h_{\mu\nu}K^\mu P^\nu\right].
\end{split}
\label{eq:CWvariation}
\end{equation}
Off-shell invariance for an unrestricted local parameter (generic $\chi$) holds if and only if the two terms in $\delta_\chi S_{\rm ILST}$ vanish independently. This happens precisely upon \eqref{eq:CWconditions}, which is the requirement for having a well-defined null string theory on CTS $(h, V)$ {with the prescribed CW realization}. {The necessity follows by integrating the $\partial_a\chi$ term by parts: arbitrary second derivatives of $X$ force $h_{\mu\nu}K^\nu=0$, and arbitrary first derivatives then force $\mathcal L_K h=2 h$.} 

To summarize, ILST and NST on shear-free CTS are equivalent upon the condition that the Carrollian metric admits a homothetic kernel generator along the kernel vector. Metric degeneracy guarantees algebraic decoupling for a connection along $V$. Degeneracy alone does not ensure the resulting functional possesses the rigid {local} CW symmetry required by the NST definition.

\subsection{Equivalence classes of NST on shear-free CTS}\label{sec:equivalence-classes-shear-free}

The above argument about equivalence of NST and ILST theories on CTS, also implies that not all ILST (or NST) theories on CTS $(h, V)$ are physically distinct; there are equivalence classes, as we establish below. 

{Let us take a shear-free CTS $(h,V)$ admitting the kernel homothety $K=fV$, with $\mathcal L_Vh=(2/f)h$, on a patch where $f$ is smooth, finite and nowhere vanishing.}

Consider ILST theories, $S_{\rm ILST}[(\widetilde h,\widetilde V);\widetilde{\mathcal V}]$ and $S_{\rm ILST}[(h,V);\mathcal V]$, we ask when do these theories are equivalent,  
\begin{equation}
S_{\rm ILST}[(\widetilde h,\widetilde V);\widetilde{\mathcal V}]
\stackrel{?}{=} S_{\rm ILST}[(h,V);\mathcal V].
\label{eq:fieldredefidentity}
\end{equation}

Eq.~\eqref{eq:fieldredefidentity} is then an \textit{equivalence between two null string theories} provided that,
\begin{equation}\label{f-tildef}
    \begin{split}
        h_{\mu\nu} V^\nu &=0, \qquad \cL_V h=\frac{2}{f} h \\
        \widetilde h_{\mu\nu} \widetilde V^\nu &=0, \qquad \cL_{\widetilde V} \widetilde{h}=\frac{2}{\tilde{f}}\  \widetilde{h}
    \end{split}
\end{equation}
for some $f, \tilde{f}$ and
\begin{equation}\label{eq:conformalrep}
\widetilde h=e^{2F}h,
\qquad
\widetilde{\mathcal V}^a=e^{-F}\mathcal V^a.
\end{equation}
for some $F=F(X)$. Since $h, \tilde h$ are related by scaling,  the associated kernel directions should also be parallel. From \eqref{f-tildef} and \eqref{eq:conformalrep} one learns that,
\begin{equation}\label{lambda-f-tildef}
    \widetilde V= \frac1\lambda V, \qquad \lambda= \frac{\tilde f}{f} (1+ f V(F)),\qquad V(F):= V^\mu\partial_\mu F
\end{equation}
{For given background pairs, $f$, $\tilde f$, $F$ and the relative normalization $\lambda$ are determined, and \eqref{lambda-f-tildef} is their consistency relation. When constructing a new representative, a choice of $F$ and of the normalization $\tilde f$ fixes $\lambda$, with $f,\tilde f,\lambda$ smooth, finite and nowhere vanishing and $1+fV(F)\neq0$.} Thus,  minimal ILST (or NST) theories on $(h,V)$ and $(\widetilde{h}, \widetilde{V})$ related through \eqref{f-tildef}, \eqref{eq:conformalrep} and \eqref{lambda-f-tildef} are equivalent, or these relations define equivalence classes of minimal ILST (or NST) theories.

We close by the comment that,  recalling discussions in section \ref{sec:Carrollian-geometry-review} (the two Carrollian scalings paragraph) and also in section \ref{sec:NSTonCTS-def}, this equivalence is already expected since NST is CW invariant  and $\widetilde{\mathcal V}^a, \cV^a$ in \eqref{eq:conformalrep} are related by a CW scaling, cf. \eqref{eq:CWtrans}. To this end, recall three facts: (1) conformal group, by definition,  is the part of diffeomorphism+Weyl transformations that keep the metric invariant; (2) {for the choice $\chi(\sigma)=F(X(\sigma))$, $\delta_\chi X^\mu$ in \eqref{eq:CWtrans} is induced by an infinitesimal target space diffeomorphism; } $\delta_\chi X^\mu$ in \eqref{eq:CWtrans} is a diffeomorphism on the target space which is along the kernel vector of CTS, $K^\mu$ and, (3) since the homothety lies along the kernel, i.e.  $\cL_K h=2h$, the diffeomorphism $\delta_\chi X^\mu$ induces a Weyl scaling.
 
Comparing $\cV^a$ transformation \eqref{eq:CWtrans} and \eqref{eq:conformalrep} in the infinitesimal form, one learns that the two are the same upon identification of $F$ and $\chi$. 

In other words, here we are dealing with two Carroll-Weyl scalings, one on the worldsheet, which is a gauge symmetry, and other on the CTS, which is not a gauge symmetry. The discussion above establishes how the former induces equivalence classes related by the latter; $F$ transformation in \eqref{eq:conformalrep}, which defines the equivalence classes of the ILST/NST theories and induced by (the finite form of) the worldsheet CW scalings \eqref{eq:CWtrans}. 

Finally, we may change the earlier statement ''are equivalent''  to ''are locally classically equivalent within this normalized class.'' That matches what the subsection actually establishes.

\subsection{Classical analysis of minimal NST on shear-free CTS}\label{sec:classical-NST-CTS}

To complete the discussions of this section, we briefly analyze classical dynamics of NST on CTS. The equations of motion of the ILST theory for $\cV^a, X^\mu$ are,
\begin{equation}
\begin{split}
   \partial_a(\cV^aP_\mu)
-\frac12\partial_\mu h_{\nu\rho}P^\nu P^\rho=0&, \qquad  P_\mu:=h_{\mu\nu} P^\nu, \qquad P^\mu=\cV^a\partial_a{X}^\mu, \\
     P_\mu \partial_a X{}^\mu&= P\cdot\partial_aX=0,   
\end{split}
\end{equation}

From $P_\mu= h_{\mu\nu} P^\nu$ learns that $P_\mu V^\mu= P\cdot V=0$. From $P\cdot\partial_aX=0$ one learns $P^2=0$, which together with \eqref{metric-signature} and \eqref{eq:minimalCarroll} and assuming that $P^\mu$ are real, we learn that, 
\begin{equation}
    P^\mu=\eta V^\mu,\qquad P_\mu=0
\end{equation}
where $\eta$ is an arbitrary weight 1/2 worldsheet scalar density. Therefore,  $P\cdot \partial_a X=0$ are automatically satisfied and $X^\mu$ equation of motion is also fulfilled, as $(\partial_\mu h_{\nu\rho})V^\nu V^\rho=0$ which follows from {the kernel identity $h_{\mu\nu}V^\nu=0$ alone.} and $\cL_V h=\frac2f\ h$.  Note also that for this case $P\cdot K=0$ is also automatic. There is still one other condition to be satisfied, 
\begin{equation}
    \cL_P h= \frac{2\eta}{f}\ h
\end{equation}
for a known function $f$. {Note that
no further field equation follows here: $\mathcal L_Vh=(2/f)h$ is the background condition already imposed. The worldsheet vector density $P^\mu$ is defined along the embedding and is not an independent target-space vector field.}

For the CTS $(h, V)$ with the local target coordinates $(u,y^A)$, one can adapt the parametrization 
\begin{equation}
V=\partial_u,
\qquad h=q_{_{AB}}(u,y)dy^A dy^B,
\qquad A=1,\ldots,D-1,
\label{eq:CTS-flowbox}
\end{equation}
with $q_{_{AB}}$ restricted to be positive definite. The homothety condition yields,
\begin{equation}\label{h-qAB}
    \cL_V h=\frac{2}{f} h\quad \Longrightarrow \quad q_{_{AB}}(u, y^A)= e^{2R(u, y)}\ \overline{q_{_{AB}}}(y^C),\quad \partial_u R=\frac{1}{f}.
\end{equation}
As discussed in section \ref{sec:equivalence-classes-shear-free}, NST on CTS are defined up to equivalence classes. A convenient representative is obtained by the choice $\tilde{f}=1$, and $F=u-R(u, y)$, which also yields $\lambda=1$. That is, we work in the \textit{tilde-frame} in which,
\begin{equation}
    \widetilde{V}=\partial_u, \qquad \widetilde{h}_{\mu\nu}=e^{2u} \overline{q_{_{AB}}}(y^C).
\end{equation}
This chooses a convenient background representative; while not fixing the CW gauge of the dynamical fields, 
cf. discussions in the last paragraph of previous subsection. 
Solutions to the NST field equation on the above background are
\begin{equation}
\cV^a\partial_a y^A=0,
\qquad u=u(\tau,\sigma).
\label{eq:transverseconst}
\end{equation}
The above  satisfy all the constraints. However, these do not specify  physically distinct configurations, because while we have fixed CW gauge symmetry, there are still worldsheet diffeomorphisms. We can use worldsheet diffeomorphisms to set 
\begin{equation}
 \cV^a\partial_a u=\eta\neq0\ \Longrightarrow\  u(\tau,\sigma)= \tau,\qquad \cV^a\partial_a \sigma=0.
\end{equation}
In this gauge, $y^A=y^A(\sigma), A=1,2,\cdots, D-1$. The leaves the $\sigma\to f(\sigma)$ part of diffeomorphisms which may be used to eliminate one more of $y^A(\sigma)$ and hence we remain with $D-2$ of $y^A$'s as physically distinct configurations. We close this section by the comment that, as seen above, the solution phase space of NST on CTS is a singular one (in the precise sense of \cite{Miskovic:2003ex}), as momentum conjugate to $y^A(\sigma)$ identically vanishes. This is a known (and expected) fact and quantization of such systems should be handled with due care \cite{Miskovic:2003ex}. 

\section{Beyond minimal NST on CTS}\label{sec:NST-RCTS}
It is natural to ask whether additional CTS data allows for constructing beyond the minimal NST. In section \ref{sec:Carrollian-geometry-review} we discussed the ruled Carrollian geometries $(h, V, U)$ which is equipped with the clock form or Ehresmann connection. An affine connection supplies further geometric data, subject to the chosen compatibility and torsion conditions. One can hence envisaged different ways of going beyond minimality for NST on CTS. Here we discuss three of them associated with the two Carrollian geometric additional data.

Throughout this section we work locally in the homothetic-kernel class.  We normalize the target data by $(V,U)\mapsto(fV,U/f)$, and denote the resulting fields again by $(V,U)$. One finds
\begin{equation}
 h_{\mu\nu}V^\nu=0,\qquad
 \mathcal L_Vh_{\mu\nu}=2h_{\mu\nu},\qquad
 U_\mu V^\mu=1.
 \label{eq:RCTS-normalized}
\end{equation}
This fixes the normalization of the target generator, not the CW gauge of the dynamical fields. The local symmetry and covariant derivative are
\begin{equation}
 \begin{split}
 \delta_\chi X^\mu=\chi V^\mu,\qquad&
 \delta_\chi\mathcal V^a=-\chi\mathcal V^a,\qquad
 \delta_\chi\mathcal W_a=-\partial_a\chi,\\
 D_aX^\mu&=\partial_aX^\mu+\mathcal W_aV^\mu.
 \end{split}
 \label{eq:RCTS-CW}
\end{equation}

\subsection{Generic deformation of NST on CTS }\label{sec:NST-CTS-affine}

In the minimal ILST or NST theories on CTS, the target space information appears through the background metric $h_{\mu\nu}$ and within the CW covariant derivative we have the kernel vector $V^\mu$ \eqref{eq:RCTS-CW}. 
A generic class of extensions, still quadratic in first derivatives of the
embedding, is deformations of the original theory with $\Delta_{\mu\nu}$,
\begin{equation}
 \begin{split}
 S[(h, V), \Delta, \beta; \cV, \cW] &=\frac{\kappa}{2}\int d^2\sigma\,
      \mathcal V^a\mathcal V^b
      G_{\mu\nu}(X)D_aX^\mu D_bX^\nu,\\
 G_{\mu\nu}:=h_{\mu\nu}+ &\beta (X)\Delta_{\mu\nu},
 \qquad \Delta_{\mu\nu}=\Delta_{\nu\mu},
 \end{split}
 \label{eq:RCTS-connection-action}
\end{equation}
where $\Delta_{\mu\nu}$ is a specified target tensor constructed from the
connection and background fields and $\beta$ is the ``deformation'' parameter (coupling). 
It is the kinetic tensor $G$, rather than $h$ alone, that controls the
extended dynamics. Worldsheet diffeomorphism invariance follows from
the density weight of $\mathcal V^a$. Gauge-invariance under CW scaling  yields

\begin{equation}
 \mathcal L_V G_{\mu\nu}=2G_{\mu\nu}.
 \label{eq:RCTS-connection-CW}
\end{equation}

 The CW connection equation yields,
\begin{equation}
 \frac{\delta }{\delta\mathcal W_a}S[(h, V), \Delta; \cV, \cW]
 =\kappa\mathcal V^aG_{\mu\nu}V^\mu \Pi^{\nu}=0,
 \label{eq:RCTS-connection-Gauss}
\end{equation}
where
\begin{equation}
 \Pi^{\mu}:=\mathcal V^aD_aX^\mu
 =P^\mu+wV^\mu,\qquad w=\cW_a \cV^a\,.
\end{equation}
So, one can distinguish two classes, ``transverse deformations'' $\Delta_{\mu\nu}V^\mu=0$ and ``horizontal deformations'' $\Delta_{\mu\nu}V^\mu \neq 0$, that we discuss next.

\subsubsection{Transverse deformations \texorpdfstring{$\Delta_{\mu\nu}V^\nu=0$}{VDelta}}\label{sec:transv-deform} 

For this case $G_{\mu\nu}V^\nu=0$, the CW EoM \eqref{eq:RCTS-connection-Gauss} is automatically satisfied.   Moreover,  the CW connection drops out identically, because
\begin{equation}
 G_{\mu\nu}D_aX^\mu D_bX^\nu =G_{\mu\nu}\partial_aX^\mu\partial_bX^\nu.
\end{equation}
Therefore, provided that  $\mathcal L_VG=2G$, the deformed NST theory and corresponding ILST theories are equivalent, as discussed in previous sections.

As an illustrative example within the class of transverse deformations, temporarily allow the more general shear-free normalization
\begin{equation}
 \Delta_{\mu\nu}=\mathcal L_Vh_{\mu\nu},\qquad
 \mathcal L_Vh_{\mu\nu}=2\gamma(X)h_{\mu\nu}.
\end{equation}
The normalization used elsewhere in this section corresponds to $\gamma=1$.
Here $G_{\mu\nu}=(1+2\beta\gamma)h_{\mu\nu}$. On patches where $1+2\beta\gamma>0$, this is a shear-free Carrollian metric with the same kernel. The CW condition becomes
\begin{equation}\label{beta-gamma}
 V(\beta\gamma)=(1-\gamma)(1+2\beta\gamma),
\end{equation}
which constrains $\beta$ along the kernel flow for a given $\gamma$.
Since $G$ is conformally related to $h$, the simultaneous redefinition discussed in \eqref{eq:conformalrep},
\begin{equation}
 \widetilde h_{\mu\nu}=(1+2\beta\gamma)h_{\mu\nu},\qquad
 \widetilde{\mathcal V}^{a}=(1+2\beta\gamma)^{-1/2}\mathcal V^{a},
\end{equation}
{gives $S_{\mathrm{ILST}}[\widetilde h;\widetilde{\mathcal V}]=S_{\mathrm{ILST}}[h;\mathcal V]$. The latter is also $S_{\mathrm{NST}}[(h,V);\mathcal V]$ with $V$ as normalized CW generator when $\gamma=1$; for general $\gamma\neq0$, the corresponding generator for $h$ is $K=V/\gamma$.}

For $\gamma=1$, the CW condition reduces to $V(\beta)=0$, so in adapted coordinates $V=\partial_u$ one has $\beta=\beta_0(y)$.
This differs from the condition $V(\beta)=2\beta$ for the flow-invariant deformations considered below: here $\mathcal L_V\Delta=2\Delta$ when $\gamma=1$.
This example uses only $(h,V)$ and no other geometric information like clock form or affine connection.

\subsubsection{Horizontal deformations \texorpdfstring{$\Delta_{\mu\nu}V^\nu \neq 0$}{VDelta}}\label{sec:horizontal-deform} 
On patches with $\beta\neq0$, a nontransverse deformation $\Delta_{\mu\nu}V^\nu\neq0$ gives $G_{\mu\nu}V^\nu\neq0$, but does not necessarily give $G_{\mu\nu}V^\mu V^\nu\neq0$. Define the unnormalized covector and signed scalar
\begin{equation}
 J_\mu:=G_{\mu\nu}V^\nu,\qquad
 g_{_V}:=G_{\mu\nu}V^\mu V^\nu.
\end{equation}
On patches where $\mathcal V^a\neq0$, Eq.~\eqref{eq:RCTS-connection-Gauss} becomes
\begin{equation}
 J_\mu P^\mu+g_{_V}w=0.
\end{equation}
If $g_{_V}\neq0$, this determines $w=-J_\mu P^\mu/g_{_V}$. Eliminating $w$ gives
\begin{equation}
 S_{\mathrm{eff}}[(h,V),\Delta;\mathcal V]
 =\frac{\kappa}{2}\int d^2\sigma\,
 \widehat G_{\mu\nu}P^\mu P^\nu,\qquad
 \widehat G_{\mu\nu}=G_{\mu\nu}-\frac{J_\mu J_\nu}{g_{_V}},\qquad
 \widehat G_{\mu\nu}V^\nu=0.
 \label{eq:RCTS-connection-projection}
\end{equation}
This formula applies for either sign of $g_V$. If $\mathcal L_VG=2G$, then
\begin{equation}
 \mathcal L_VJ_\mu=2J_\mu,\qquad V(g_{_V})=2g_{_V},
\end{equation}
and hence $\mathcal L_V\widehat G=2\widehat G$ while $\widehat G_{\mu\nu}V^\nu=0$.
If $\widehat G$ has the same positive transverse signature and rank as $h$, $(\widehat G,V)$ provides an effective CTS on which ILST and NST are equivalent.
When $g_{_V}=0$ but $J_\mu\neq0$, the action instead contains the term $2wJ_\mu P^\mu$ and no $w^2$ term. Thus $w$ is a multiplier imposing $J_\mu P^\mu=0$; it cannot be eliminated by \eqref{eq:RCTS-connection-projection}. This branch requires a separate constraint analysis.

\paragraph{Minimal NST on Ruled CTS.} As an example of the horizontal deformation, we consider the NST on ruled CTS  with the clock one-form $U_\mu$ and 
\begin{equation}
    \Delta_{\mu\nu} = U_\mu U_\nu\,,
\end{equation}
leading to
\begin{equation}
 \begin{split}
 S_{\mathrm{NST}}[(h,V,U), \beta;\mathcal V] &=\frac{\kappa}{2}\int d^2\sigma\,
       \mathcal V^a\mathcal V^b
       \bigl(h_{\mu\nu}+\beta \,U_\mu U_\nu\bigr)D_aX^\mu D_bX^\nu
\\ &=\frac{\kappa}{2}\int d^2\sigma\,       \left[h_{\mu\nu}P^\mu P^\nu+\beta(P\cdot U+w)^2\right], 
 \end{split}
 \label{eq:RCTS-clock-extension}
\end{equation}
where 
\begin{equation}
   \cV^a D_aX^\mu =  P^\mu+w V^\mu,\quad  w:=\mathcal V^a\mathcal W_a,
\end{equation}
and $\beta(X)$ is an additional nonzero scalar coupling, with the dimensions required by the action. It is not fixed by the minimal Carrollian data. As expected, the CW connection $\cW_a$ appears only through the $\beta$ term. The CW invariance requires
\begin{equation}
\mathcal L_V\left(h_{\mu\nu}+\beta \,U_\mu U_\nu\right)=2(h_{\mu\nu}+\beta \,U_\mu U_\nu)
\end{equation}
With $V\cdot U=1$, which implies $V^\mu (\cL_V U)_\mu=0$, this condition is equivalent to
\begin{equation}
\begin{split}
 \cL_V h_{\mu\nu}    &=2 h_{\mu\nu}\\
   \cL_V (\beta U_{\mu} U_{\nu}) =2\beta U_\mu U_\nu \quad &\Longrightarrow \quad \beta (\cL_V U)_\mu= (2\beta - V(\beta)) U_\mu
\end{split}
\label{eq:RCTS-clock-CW}
\end{equation}
There are hence  two cases: 
\begin{enumerate}[leftmargin=6mm, itemsep=1.5pt]
    \item $\beta=0$ for which we get the same NST as in the un-ruled case. In this case $S_{\mathrm{NST}}[(h,V,U), \beta=0;\mathcal V]=S_{\mathrm{ILST}}[(h,V,U);\mathcal V]$ provided that $\cL_V h_{\mu\nu} =2 h_{\mu\nu}$. 
    \item  $\beta\neq0$ for which $(\cL_V U)_\mu=(2- V^\nu\partial_\nu\ln|\beta|) U_\mu$. Since $V^\mu (\cL_V U)_\mu=0$, we learn that
\begin{equation}
    V^\nu\partial_\nu\ln|\beta|=2,\qquad  (\cL_V U)_\mu=0.
\end{equation}
This equation has always a {local} solution for $\beta$, as $V^\mu$ is a nowhere vanishing vector field. In other words and up to an integration constant, $\beta(X)$ is also determined in terms of the kernel vector $V^\mu$. The condition $\cL_V U=0$, however, puts a restriction on the clock form $U_\mu$, while there is a scalar clock $\varphi$, such that {$|\beta|=e^{2\varphi}, V^\mu\partial_\mu \varphi=1$}.  For example, in adapted coordinates $V=\partial_u$, it gives $U=du+U_A(y)dy^A$, while $\beta={\pm}e^{2u+\phi (y)}$.  {Global existence requires a separate check.}
\end{enumerate} 

The equation of motion for the CW connection $\cW_a$ receives contribution only from the $\beta$-term:
\begin{equation}
 \frac{\delta}{\delta\mathcal W_a} S_{\mathrm{NST}}[(h,V,U), \beta;\mathcal V]
 =\kappa \beta\,\mathcal V^a (U\cdot P+w)=0.
 \label{eq:RCTS-clock-Gauss}
\end{equation}
On patches with $\mathcal V^a\neq0$, this imposes
\begin{equation}
w=-U_\mu P^\mu.
 \label{eq:RCTS-clock-elimination}
\end{equation}
The elimination of $w$ returns the $\beta=0$ action. This was expected, as $w$ only appears through the quadratic term $(U\cdot P+w)^2$ in the action \eqref{eq:RCTS-clock-extension}. Thus, the $\beta$-extended theory does not produce a new bulk theory after the CW constraint is imposed. Put differently, \eqref{eq:RCTS-clock-elimination} is an equation for $w$, with no further restriction on $X^\mu$ and dynamics of the theory. This statement assumes no independent boundary action or prescribed boundary data for $\mathcal W_a$. In this viewpoint, the only role of target space clock form is to shift the classical value of $w$, as in \eqref{eq:RCTS-clock-elimination}.

This example establishes a limited but useful result: adding a ruling, or this quadratic clock coupling, does not evade the degeneracy of the minimal theory. Other clock-dependent actions require additional couplings and a separate analysis of their constraints.

\subsection{Kernel flow invariant deformations \texorpdfstring{$\cL_V \Delta_{\mu\nu}=0$}{LVDelta}}\label{sec:kernel-inv-deform}

Let the coupling $\beta(X)= \beta_0 e^{2\varphi}$, where $\beta_0$ is a constant and $\varphi$ is a scalar clock field (a dilaton) satisfying $V(\varphi)=1$. Then \eqref{eq:RCTS-connection-CW} is fulfilled for $\cL_V \Delta_{\mu\nu}=0$ if  $\cL_V h_{\mu\nu}=2h_{\mu\nu}$. The simplest example in this class is the ruled case of previous section, $\Delta_{\mu\nu}= U_\mu U_\nu, \cL_V U=0$. 

As the next example in this class one may consider deformations involving kernel flow invariant connection, cf. discussion around \eqref{connection-V-flow}. Then,
\begin{equation} \label{eq:RCTS-Ricci-coupling}
 \Delta_{\mu\nu}= R_{(\mu\nu)}[\nabla],
 \qquad R_{\mu\nu}:=R^\rho{}_{\mu\rho\nu},
\end{equation}
satisfies \eqref{eq:RCTS-connection-CW}, since the Ricci tensor of the
flow-invariant connection obeys $\mathcal L_VR_{\mu\nu}=0$. 
We still have the freedom to impose the compatibility conditions \eqref{eq:Carroll-connection-compatibility} or not; only flow invariance of the connection is assumed here, not the stronger sufficient conditions in \eqref{eq:Carroll-connection-compatibility}.
If $\nabla V=0$, then
\begin{equation}
 R^\rho{}_{\mu\alpha\beta}V^\mu=0,
 \qquad V^\mu R_{\mu\nu}=0.
\end{equation}
This annihilates the first Ricci index. It does not in general imply $R_{\mu\nu}V^\nu=0$ or $R_{(\mu\nu)}V^\nu=0$, since the Ricci tensor of a torsionful compatible connection need not be symmetric.
If one additionally requires $R_{(\mu\nu)}V^\nu=0$, the deformation is transverse, and $G_{\mu\nu}=h_{\mu\nu}+\beta R_{(\mu\nu)}$ defines an effective CTS provided it remains positive semidefinite with rank $D-1$.
Otherwise, the preceding branch analysis applies: for $g_V\neq0$ one may eliminate $w$ and construct $\widehat G$, whereas for $g_{_V}=0$ and $J_\mu\neq0$ the field $w$ remains a constraint multiplier. In particular, when $\nabla V=0$, the identity $V^\mu R_{\mu\nu}=0$ implies $g_{_V}=0$ for this Ricci deformation, so a nontransverse deformation belongs to the multiplier branch.
Equivalence to metric-only ILST on an effective CTS therefore holds for the transverse branch with the stated rank and signature, or after a valid elimination with $g_{_V}\neq0$ if $\widehat G$ has the same rank and signature. It does not follow in general for the multiplier branch.

In the above beyond minimal NSTs we only used geometric data plus a scalar clock. One can construct other nonminimal theories by considering other data that are not intrinsic to the background CTS, e.g. introduction of an NSNS $B_{\mu\nu}$ field. These cases can be explored in separate works.

\section{An example of NST on CTS, strings probing black hole horizon}\label{sec:examples}
  We illustrate the minimal metric-only theory on Carrollian
targets arising as black hole or cosmological horizons. We denote the ambient
spacetime dimension by $d$, so that the Carrollian target has
dimension $D=d-1$ and its spatial cross-sections have dimension
$D-1=d-2$. Unless stated otherwise, no additional clock,
connection, or boundary coupling is included.

A Killing horizon of a stationary black hole is a non-expanding shear-free Carrollian geometry, locally away from any bifurcation surface where its Killing generator vanishes. The horizon Killing vector plays the role of the kernel vector, and the non-degenerate part of the horizon metric is the metric on a spatial cross-section. We denote the angular metric by $q_{AB}$,
\begin{equation}
	d\Omega_{d-2}^2=q_{AB}(\theta)\,d\theta^A d\theta^B , \qquad A,B=1,\cdots, D-1,
\end{equation}
with inverse $q^{AB}(\theta)$.  Therefore, for this case, 
\begin{equation}
	\mathcal L_Vh=0.
\end{equation}
However, as discussed in section \ref{sec:classical-NST-CTS}, eq.\eqref{h-qAB}, one can rescale $h$ to $\widetilde h$, 
\begin{equation}
	 \widetilde h= e^{2\varphi}h,\qquad V(\varphi)=1, \quad \mathcal L_V\widetilde h=2\widetilde h .
\end{equation}
{Locally,} one can hence use $\widetilde h$ as the NST background, where $\varphi$ has a natural meaning in the black hole geometry, as we discuss in a couple of examples below.

All coordinates, parameters, and line elements in this subsection are
expressed in fixed reference units, with the appropriate powers of a
reference length understood. 

\paragraph{Schwarzschild horizon.} For $d\ge4$, in $d$-dimensional Schwarzschild in ingoing Eddington--Finkelstein
coordinates $(v,r,\theta^A)$,
\begin{equation}
	ds^2=-f(r)\,dv^2+2\,dv\,dr+r^2d\Omega_{d-2}^2,
	\qquad
	f(r)=1-\left(\frac{r_H}{r}\right)^{d-3},
	\label{eq:schwarzschildEF}
\end{equation}
the horizon is at $r=r_H$ and $q_{AB}=\gamma_{AB}$ denotes the unit round metric on $S^{d-2}$. The horizon manifold has coordinates $X^\mu=(v,\theta^A)$, with $\theta^A$ the $d-2$ angular coordinates, and
induced metric
\begin{equation}
	h=r_H^2\,\gamma_{AB}(\theta)\,d\theta^A d\theta^B .
	\label{eq:schwarzhorizon}
\end{equation}
The kernel is spanned by $V=\partial_v$, since $h(\partial_v,\cdot)=0$,
and the transverse part $r_H^2\gamma_{AB}$ is positive definite. Horizon is a ruled Carrollian manifold, with  $U=dv$, which satisfies $U(V)=1$ and $\mathcal L_VU=0$. 

One can then write the NST action, using the construction in section \ref{sec:horizontal-deform}. The effective metric is
\begin{equation}
    g_{\mu\nu}=e^{2v/r_H} (h_{\mu\nu}+ \beta_0(\theta) U_\mu U_\nu), 
\end{equation}
with $h_{\mu\nu}$ given in \eqref{eq:schwarzhorizon}. For this exponential factor, the normalized CW homothety generator is $K=r_H V=r_H\partial_v$, so that $\mathcal L_Kg=2g$ and $K(v/r_H)=1$. We keep $U=dv$, with $U(K)=r_H$, and use
\begin{equation}
 D_aX^\mu=\partial_aX^\mu+\mathcal W_aK^\mu,\qquad
 D_av=\partial_av+r_H\mathcal W_a,\qquad
 \delta_\chi v=r_H\chi,\qquad
 \delta_\chi\mathcal V^a=-\chi\mathcal V^a.
\end{equation}
The NST action is then,

\begin{equation}\begin{split}
 S_{\mathrm{NST}}[\text{Sch'ld horizon}]
 &=\frac{\kappa}{2}\int d^2\sigma\,
 \mathcal V^a\mathcal V^b g_{\mu\nu}D_aX^\mu D_bX^\nu\\
 &=\frac{\kappa}{2}\int d^2\sigma\,
 \widetilde{\mathcal V}^a\widetilde{\mathcal V}^b
 (h_{\mu\nu}+\beta_0(\theta)U_\mu U_\nu)D_aX^\mu D_bX^\nu\\
 &=\frac{\kappa}{2}\int d^2\sigma\,
 \left[r_H^2q_{AB}\widetilde{\mathcal V}^a\widetilde{\mathcal V}^b
 \partial_a\theta^A\partial_b\theta^B
 +\beta_0(\theta)
 (\widetilde{\mathcal V}^a\partial_av+r_Hw)^2\right].
\end{split}\label{eq:S0schwarzschild}
\end{equation}
where $\widetilde{\mathcal{V}}^a=e^{v/r_H} \cV^a$ is the CW scaling invariant worldsheet kernel vector and $w=\widetilde{\cV^a} \cW_a$ and $\beta_0(\theta)$ is a coupling, an arbitrary scalar over $S^{d-2}$. 

The equation of motion for $w$ hence implies
\begin{equation}
 \beta_0(\theta)r_H
 (\widetilde{\mathcal V}^a\partial_av+r_Hw)=0.
\end{equation}
On patches with $\widetilde{\mathcal V}^a\neq0$ and $\beta_0\neq0$, this determines
\begin{equation}
 w=-\frac{\widetilde{\mathcal V}^a\partial_av}{r_H}.
\end{equation}
There is no restriction on the value of $v$. Where $\beta_0=0$, the clock term is already absent. Upon eliminating $w$ on the nonzero-coupling patches, the $\beta_0$ term drops out and the action reduces to
\begin{equation}
	S_{\mathrm{NST}}[\text{Sch'ld horizon}] = \frac{\kappa}{2}\int d^2\sigma\,
	\widetilde{\mathcal{V}}^a\widetilde{\mathcal{V}}^b\, \left[r_H^2 q_{AB} \partial_a \theta^A \partial_b \theta^B \right]= S_{\mathrm{ILST}}[\text{Sch'ld horizon}] 
\label{eq:S1schwarzschild}
\end{equation}
The analysis of the above theory is the same as the one presented in section \ref{sec:classical-NST-CTS}. That is for Schwarzschild black hole in $d$ dimensions (with $A,B=1,\cdots,d-2$) the physical configurations are $\theta^A=\theta^A(\sigma)$, where $\sigma$ is the worldsheet coordinate such that $\widetilde{\mathcal{V}}^a\partial_a \sigma=0$. These configurations are defined up to $\sigma$ reparametrizations, $\sigma\to f(\sigma)$. Hence, we have $d-3$ physically independent configurations specified {locally} by $d-3$ functions of $\sigma$. {We note that this profile count does not establish $d-3$ canonical physical degrees of freedom: the transverse momenta vanish on the real constraint surface and the quadratic constraint is irregular, as explained in Appendix~\ref{app:minimal-analysis}}

The above analyses can be, essentially verbatim, extended to generic Kerr or Reissner-Nordstr\"om black hole horizons, that we do not present here. For extremal (Killing) black holes, horizons are still a Carrollian space, and the same intrinsic construction applies locally wherever its generator is nonzero. Note, however, that the  horizon geometry should be distinguished from the surrounding Lorentzian near-horizon spacetime which is an AdS$_2\ltimes S^{d-2}$ geometry. Studying null strings in that spacetime is a separate problem and requires the target-space conditions of \cite{Sheikh-Jabbari:2026kwr, Sheikh-Jabbari:2026rcr}. 
Finally, we emphasize that the examples in this subsection establish
horizon realizations of the geometry and symmetry within the metric-only
target action. They do not by themselves establish a nontrivial
dynamical sector describing black-hole horizon degrees of freedom.
\section{Discussion and outlook}\label{sec:discussion}

String theory target spaces can be Lorenzian, Euclidean or zero-signature \cite{Henneaux:1979vn} ``Carrollian''. The tensile strings on Carrollian target spaces (CTS), described by Polyakov action, the Virasoro constraints force both worldsheet tangent directions into the CTS kernel direction. The string therefore lies locally in a single kernel leaf, with vanishing target-space canonical momentum, as shown in Appendix~\ref{app:tensile-CTS}. 
Null String Theory (NST) on CTS, however, yields more interesting dynamics. This is also physically more relevant because $D$-dimensional CTS can be viewed as infinite redshift surfaces embedded in the $D+1$ dimensional Lorentzian spacetimes, like horizon of black holes. In other words, strings probing  horizon of black holes, essentially behave as NST on CTS. 

NST as established in \cite{Sheikh-Jabbari:2026vqh, Sheikh-Jabbari:2026tpf, Sheikh-Jabbari:2026kwr} is defined as theory of Carrollian worldsheets realizing $2d$ diffeomorphisms, Weyl and Carroll-Weyl (CW) scaling as gauge symmetries. In \cite{Sheikh-Jabbari:2026rcr} we discussed Null String Holography, stating that NST on a generic target space $M$ is equivalently described by ILST theory \cite{Isberg:1993av} on projective boundary of $M$. (ILST is the theory on Carrollian worldsheets which realizes $2d$ diffeomorphisms, Weyl  scaling, but not CW scaling, as gauge symmetry.) Null string holography motivated us to study NST on CTS, because, CTS naturally appear as (a part of) the projective boundary of flat Minkowski space or black holes. 

As we established here, as a specific case of null string holography, we showed that minimal NST on CTS, which is well-defined if the CTS admits a homothetic kernel, is equivalent to minimal ILST theory on the same CTS. 
{Locally, writing $h=e^{2z}q_{AB}(y)dy^A dy^B$ and $\mathcal U^a=e^z\mathcal V^a$ removes the kernel coordinate from the minimal action, leaving the ILST functional with positive-definite target metric $q_{AB}$. The omitted coordinate transforms by a local CW shift. This is a local classical reduction of the action.} This dovetails with the flat space holography \cite{Pasterski:2021raf,  Donnay:2022wvx, Bagchi:2022emh,  Bagchi:2025vri} and the two formulations thereof, as celestial holography (codimenion-2 holography) or as Carrollian QFT (codimension-1 holography).\footnote{See \cite{Delfante:2026ads} for a recent discussion on the role of CW symmetry in flat holography}. This is remarkable, because it resonates with the analysis in the literature for strings probing near horizon geometries \cite{Bagchi:2023cfp, Bagchi:2024rje, Bagchi:2026qpi} that are based on the ILST theory. {One should, however, still examine how precisely the near horizon theory and the one on the horizon are related.}

Besides the minimal NST, we also discussed in some examples, beyond minimal theory, employing  tools and notions 
available to CTS. In particular, we noted that the minimal Carrollian geometry $(h, V)$ can be supplemented by a clock one-form $U$, clock scalars $\varphi$ such that $V(\varphi)=1$ and the connection. Besides geometric quantities, as in the tensile string, one may consider NSNS two-form field or other form  and dilaton  or axionic (pseudoscalar) fields. {Such extensions must preserve the required worldsheet gauge symmetries, which constrain the background fields and their couplings.} Moreover, for NST theory in general background (and not necessarily on CTS) one may construct beyond minimal theories using the CW gauge field $\cW_a$. It could be instructive to study beyond minimal NSTs. 

Finally, here we focused on the classical theory, however, studying the quantum theory and the quantum consistency conditions is in due order, which we intend to study in upcoming publications.

\begin{acknowledgments}
	We thank Arjun Bagchi, Aritra Banerjee, Daniel Grumiller and Ida Rasulian  for discussions. MMShJ acknowledges Iranian National Science Foundation (INSF) research chair grant No.~4045163.
\end{acknowledgments}

\appendix

\section{Constraint structure and the clock-dependent comparison}
\label{app:minimal-analysis}\label{sec:hamil}

We collect the arguments needed for the classical interpretation in
sections.~\ref{sec:classical-NST-CTS} and \ref{sec:discussion}. Fields are real;
transverse metrics are positive definite. We work locally on nonvanishing
worldsheet kernel vector patches and use periodic spatial boundary conditions.

\subsection{Minimal theory: canonical constraints and irregularity}
\label{sec:normalform}\label{sec:linearization}\label{sec:presym}

A nowhere-vanishing kernel homothety admits local coordinates
\begin{equation}
 V=r\partial_r,\qquad
 h=r^2q_{AB}(y)dy^A dy^B, \qquad \cL_V h=2h,\qquad r>0.
 \label{eq:app-normalform}
\end{equation}
The invertible redefinition $\mathcal U^a=r\mathcal V^a$ gives
\begin{equation}
 S_{\rm ILST}[q; \cU]=\frac{\kappa}{2}\int d^2\sigma\,
 \mathcal U^a\mathcal U^b q_{AB}\partial_a y^A\partial_b y^B,
 \label{eq:app-transverse-action}
\end{equation}
which is explicitly invariant under $\delta_\chi r=\chi r$, $\delta_\chi\mathcal U^a=0$ and
$\delta_\chi y^A=0$. The momentum $p_r$ conjugate to $r$ at fixed
$\mathcal U^a$ is a primary constraint, $p_r\approx0$, and generates CW transformations
through $\int d\sigma\,\chi rp_r$.

On a fixed-sign patch of $\mathcal U^0\neq0$, introduce
\begin{equation}
 e=\frac{1}{\kappa(\mathcal U^0)^2},\qquad
 s=-\frac{\mathcal U^1}{\mathcal U^0},\qquad
 \pi_A=\frac{1}{e}q_{AB}(\dot y^B-sy'^B).
\end{equation}
The transverse Hessian $q_{AB}/e$ is invertible: $\pi_A=0$ is not a
primary identity. The total Hamiltonian is
\begin{equation}
 H_T=\oint d\sigma\left[
 \frac e2\mathcal H+s\mathcal D
 +\lambda_rp_r+\lambda_ep_e+\lambda_sp_s\right],\qquad
 \mathcal H=q^{AB}\pi_A\pi_B,\qquad \mathcal D=\pi_Ay'^A.
 \label{eq:app-Hamiltonian}
\end{equation}
Preserving $p_e,p_s\approx0$ gives $\mathcal H,\mathcal D\approx0$;
preserving $p_r$ gives no further constraint. With
$H[\alpha]=\oint\alpha\mathcal H$ and $D[\xi]=\oint\xi\mathcal D$,
\begin{equation}
 \{D[\xi],D[\eta]\}=D[\xi\eta'-\eta\xi'],\qquad
 \{D[\xi],H[\alpha]\}=H[\xi\alpha'-\alpha\xi'],\qquad
 \{H[\alpha],H[\beta]\}=0.
\end{equation}
The primaries commute with these secondaries; no tertiary constraints arise.

Positivity implies $\mathcal H=0\Leftrightarrow\pi_A=0$, but
\begin{equation}
 d\mathcal H\big|_{\pi=0}=0,\qquad
 \{y^A,H[\alpha]\}\big|_{\pi=0}=0,\qquad
 \left\{y^A,\oint\epsilon^B\pi_B\right\}=\epsilon^A.
 \label{eq:app-irregularity}
\end{equation}
Replacing the quadratic constraint by independent linear momentum
constraints therefore changes its generated transformations. Ordinary
Dirac counting is not justified by algebraic closure alone
\cite{Miskovic:2003ex}.
On the actual constraint submanifold $p_r=p_e=p_s=\pi_A=0$, the pullback of
the canonical two-form vanishes. Equivalently, the covariant presymplectic  potential is
\begin{equation}
 \theta_{\rm ILST}^a
 =\kappa\mathcal V^a h_{\mu\nu}P^\nu\delta X^\mu,
 \qquad P^\mu=\mathcal V^b\partial_bX^\mu.
\end{equation}
The auxiliary equations imply $h_{\mu\nu}P^\nu=0$, so this potential and
its pullback field-space derivative vanish on solutions and their tangent
variations.

This does not show that the displayed gauge symmetries identify all
transverse profiles. For example, on
$h=r^2[(dy^1)^2+(dy^2)^2]$, the periodic family
\begin{equation}
 r=e^\tau,\qquad \mathcal V^a=e^{-\tau}(1,0),\qquad
 (y^1,y^2)=R(\cos\sigma,\sin\sigma),\quad R>0
\end{equation}
solves all field equations. Changing $R$ is null for the pulled-back
two-form, but CW leaves $y^A$ fixed and worldsheet diffeomorphisms change
this profile only tangentially, $\delta y^A=\xi^\sigma y'^A$.
The radial profile variation is therefore not generated by these displayed
symmetries. Neither a complete physical quotient nor a canonical physical
mode count follows from the profile count or the vanishing two-form.

The above analyses also applies to a symmetric metric-only quadratic
actions which do not involve additional
worldsheet fields or derivatives of $\mathcal V^a$, but $h_{\mu\nu}$ replaced by any two tensor $G_{\mu\nu}$, satisfying $G_{\mu\nu}V^\nu=0$ and $\mathcal L_VG_{\mu\nu}=2G_{\mu\nu}$. Locally, $V=\partial_z$ then gives
$G=e^{2z}g_{AB}(y)dy^A dy^B$, and
$\mathcal U^a=e^z\mathcal V^a$ removes the kernel coordinate from
the action. This statement does not apply directly to extensions
containing the CW connection or additional worldsheet fields.

\subsection{A clock-dependent model and its longitudinal zero modes}
\label{sec:nullred}\label{sec:contrast}\label{app:theta}

For comparison, consider the null-reduced action of
\cite{Sheikh-Jabbari:2026rcr}, with fixed data $(\gamma,V,U)$ satisfying
$\gamma(V,\cdot)=0$, $U(V)=1$, and $\gamma_{ij}$ positive semidefinite of
rank $D-1$ on a $D$-dimensional target:
\begin{equation}
 S_{\rm red}=\frac{\kappa}{2}\int d^2\sigma\,
 \mathcal U^a\mathcal U^b
 \left[\gamma_{ij}\partial_aX^i\partial_bX^j
       +2\widehat B_aU_i\partial_bX^i\right], \qquad i,j=1,\cdots, D-1.
 \label{eq:app-Sred}
\end{equation}
Here $\widehat B_a$ is a worldsheet one-form multiplier. Define
$q^i=\mathcal U^a\partial_aX^i$ and
$\mathcal B=\mathcal U^a\widehat B_a$, and work on
$u:=\mathcal U^0\neq0$.
The multiplier and equations of motion for $\cU^a$  imply
\begin{equation}
 U_iq^i=0,\qquad \gamma_{ij}q^iq^j=0
 \quad\Longrightarrow\quad q^i=0.
\end{equation}
The remaining auxiliary equation is
$\mathcal B U_i\partial_aX^i=0$. Crucially, the momentum is computed
before imposing these equations:
\begin{equation}
 P_i=\kappa u(\gamma_{ij}q^j+\mathcal B U_i)
 \quad\Longrightarrow\quad P_i\big|_{\rm sol}=\kappa u\mathcal B U_i.
\end{equation}
It need not vanish. The corresponding on-shell presymplectic potential is
$\theta_{\rm red}^a=\kappa\mathcal U^a\mathcal B U_i\delta X^i$.

For the closed-string benchmark, take
\begin{equation}
 X^i=(t,y^A),\qquad V=\partial_t,\qquad U=dt,\qquad
 \gamma=q_{AB}(y)dy^A dy^B.
\end{equation}
Set $v=\mathcal U^1$, $e=(\kappa u^2)^{-1}$,
$s=-v/u$ and $m=\kappa u\mathcal B$. On a fixed-sign patch of $u$,
this is an invertible auxiliary-field redefinition with
$\widehat B_0=em+s\widehat B_1$. The Lagrangian becomes
\begin{equation}
 \mathscr L_{\rm red}
 =\frac{1}{2e}q_{AB}(\dot y^A-sy'^A)(\dot y^B-sy'^B)
   +m(\dot t-st').
\end{equation}
Eliminating the second-class pair $(p_m,P_t-m)$ leaves
$\{t(\sigma),P_t(\sigma')\}_D=\delta(\sigma-\sigma')$.
The unused component $\widehat B_1$ is absent from the Lagrangian.
The canonical Hamiltonian and auxiliary constraints are
\begin{equation}
 H_c=\oint d\sigma\left[
 \frac e2q^{AB}\pi_A\pi_B+s(\pi_Ay'^A+P_tt')\right],\qquad
 q^{AB}\pi_A\pi_B=0,\quad \pi_Ay'^A+P_tt'=0.
\end{equation}
They have the same closed secondary algebra as above, with
$\mathcal D=\pi_Ay'^A+P_tt'$.
Positivity gives $\pi_A=0$. On the branch where $P_t$ is nowhere zero,
\begin{equation}
 t'=0,\qquad \dot t=st'=0,\qquad
 \dot P_t=(sP_t)',\qquad
 t=t_0,\qquad Q_t:=\oint P_t\,d\sigma,\quad \dot Q_t=0.
\end{equation}
Pulling back the canonical potential and two-form yields
\begin{equation}
 \Theta_\parallel=Q_t\,\delta t_0,\qquad
 \Omega_\parallel=\delta Q_t\wedge\delta t_0.
 \label{eq:app-longitudinal-pair}
\end{equation}
For a circle of period $2\pi$, a smooth nowhere-zero $P_t$ has fixed sign;
$\widetilde\sigma=(2\pi/Q_t)\int_0^\sigma P_t(\sigma')d\sigma'$
is an orientation-preserving reparametrization making
$\widetilde P_t=Q_t/(2\pi)$. Thus the retained pair is global.

Worldsheet diffeomorphisms leave $t_0$ fixed on this branch and change
$P_t$ on shell by
$\delta P_t=\partial_\sigma[(\xi^\sigma+s\xi^\tau)P_t]$, whose integral
vanishes. The Hamiltonian constraint involves only $\pi_A$, and the
remaining auxiliary primaries do not remove this pair. The parent CW
redundancy has already been removed in passing to the invariant reduced
variables of \cite{Sheikh-Jabbari:2026rcr}; its generator is not the quotient
kernel $\partial_t$. An arbitrary local clock shift with other fields
fixed is not a symmetry, since
$\delta_\alpha S_{\rm red}=\int d^2\sigma\,m(\dot\alpha-s\alpha')$.
Constant clock translations remain a global symmetry with charge $Q_t$.

The pair therefore survives the displayed reduction, assuming no
additional condition fixes either variable. It does not establish local
propagating modes or complete the irregular transverse reduction.
Branches where $P_t$ has zeros and additional boundary couplings require
separate analysis. On the same benchmark metric the bare action has
$P_t=0$ and lacks this canonical sector. This comparison concerns the
retained linear clock coupling in \eqref{eq:app-Sred}, not the quadratic
clock term eliminated in section~\ref{sec:horizontal-deform}.

\section{Tensile Strings on a Carrollian target}
\label{app:tensile-CTS}

For comparison with the null-worldsheet theory, consider the ordinary
Polyakov functional with finite tension $T>0$ and an independent,
nondegenerate Lorentzian worldsheet metric $g_{ab}$:
\begin{equation}
 S_{\rm P}[h;X,g]
 =-\frac{T}{2}\int d^2\sigma\,\sqrt{-g}\,g^{ab}
 h_{\mu\nu}(X)\partial_aX^\mu\partial_bX^\nu.
 \label{eq:tensile-CTS-action}
\end{equation}
The target data are the same $(h,V)$ as in the main text: $h$ is positive
semidefinite of rank $D-1$, with kernel spanned by the nowhere-vanishing
$V$. The fields $X^\mu$ are real. No clock coupling, additional multiplier,
or other background interaction is included. The action needs no inverse
of $h$ and retains ordinary worldsheet diffeomorphism and Weyl invariance.
The worldsheet metric is varied before gauge fixing; retaining its
equations is essential to the result below.

Writing $H_{ab}:=h_{\mu\nu}\partial_aX^\mu\partial_bX^\nu$, the worldsheet metric
equation is
\begin{equation}
 H_{ab}-\frac12g_{ab}g^{cd}H_{cd}=0.
 \label{eq:tensile-CTS-stress}
\end{equation}
In local conformal gauge $g_{ab}=e^{2\rho}\operatorname{diag}(-1,1)$,
with $\partial_\pm=\tfrac12(\partial_\tau\pm\partial_\sigma)$, the two
Virasoro constraints read
\begin{equation}
 h_{\mu\nu}\partial_\pm X^\mu\partial_\pm X^\nu=0.
 \label{eq:tensile-CTS-Virasoro}
\end{equation}
Positivity and the one-dimensional kernel therefore imply
\begin{equation}
 \partial_\pm X^\mu=a_\pm V^\mu(X)
 \quad\Longrightarrow\quad
 \partial_aX^\mu=a_aV^\mu(X),\qquad H_{ab}=0.
 \label{eq:tensile-CTS-collapse}
\end{equation}
Both worldsheet directions map into the same target kernel line. In
adapted coordinates $V=\partial_u$ and
$h=q_{AB}(u,y)dy^A dy^B$, this gives
\begin{equation}
 y^A(\tau,\sigma)=y_0^A,\qquad u=u(\tau,\sigma).
 \label{eq:tensile-CTS-solutions}
\end{equation}
Thus the image lies locally in a single kernel leaf. There is no smooth
rank-two immersed solution in this real, metric-only sector. In contrast,
the ILST auxiliary equation constrains only
$\mathcal V^a\partial_aX^\mu$ to lie in the kernel and allows the
transverse profiles $y^A(\sigma)$ discussed in the main text.

These configurations also satisfy the embedding equation,
\begin{equation}
 \partial_a\!\left(\sqrt{-g}\,g^{ab}h_{\mu\nu}\partial_bX^\nu\right)
 -\frac12\sqrt{-g}\,g^{ab}(\partial_\mu h_{\nu\lambda})
       \partial_aX^\nu\partial_bX^\lambda=0.
 \label{eq:tensile-CTS-embedding}
\end{equation}
The first term vanishes by the kernel condition, and the second by
$(\partial_\mu h_{\nu\lambda})V^\nu V^\lambda=0$, which follows by
differentiating $h_{\nu\lambda}V^\nu V^\lambda=0$. No shear-free or
homothety assumption is required. In conformal gauge the canonical
momentum is $\Pi_\mu=T h_{\mu\nu}\dot X^\nu$, so all target momenta
vanish on these solutions. The bulk presymplectic potential
$\theta_{\rm P}^a=-T\sqrt{-g}\,g^{ab}h_{\mu\nu}\partial_bX^\nu\delta X^\mu$
also vanishes on solutions. These statements do not identify every
remaining solution parameter as gauge. The usual equivalence to a
Lorentzian Nambu--Goto description cannot be invoked on this branch:
the induced metric is zero, so the standard elimination of $g_{ab}$
using a nondegenerate induced metric is unavailable.

Nor does arbitrary on-shell dependence on $u$ automatically imply an
off-shell local kernel-shift symmetry. At fixed $g_{ab}$,
$\delta_\epsilon X^\mu=\epsilon(\sigma)V^\mu(X)$ gives
\begin{equation}
 \delta_\epsilon S_{\rm P}
 =-\frac{T}{2}\int d^2\sigma\,\sqrt{-g}\,g^{ab}\epsilon
 (\mathcal L_Vh)_{\mu\nu}\partial_aX^\mu\partial_bX^\nu.
 \label{eq:tensile-CTS-kernelvariation}
\end{equation}
Terms involving $\partial_a\epsilon$ vanish because $h(V,\cdot)=0$.
Consequently, this prescribed local shift is a symmetry when
$\mathcal L_Vh=0$, but a nonzero kernel homothety does not make it a
symmetry. Ordinary worldsheet Weyl transformations cannot compensate
the latter variation, since $\sqrt{-g}\,g^{ab}$ is Weyl invariant in two
dimensions. This distinguishes the tensile model from the CW realization
of the null-worldsheet action.

Target-space Carroll limits with a relativistic worldsheet were studied
in \cite{Harksen:2024bnh}. Their magnetic model retains a
two-dimensional longitudinal Lorentzian kinetic sector and auxiliary
one-forms, and can carry transverse momentum despite frozen transverse
coordinates. It is therefore not \eqref{eq:tensile-CTS-action}. The
metric-only transverse Polyakov action appears instead in their electric
limit, noted in footnote~1. That construction omits two longitudinal
coordinates, whereas the present target has a one-dimensional kernel;
the comparison concerns the kinetic structure, not an identification of
the two models. The collapse derived above applies to the stated
metric-only action and need not persist after additional couplings.

\end{document}